\documentclass[preprint,12pt]{elsarticle}

\usepackage[breaklinks=true,colorlinks=true,backref,pagebackref]{hyperref}

\usepackage{epsf}
\usepackage{color}
\usepackage{dcolumn}
\usepackage{bm}
\everymath{\displaystyle}

\def\a{\alpha}
\def\b{\beta}
\def\g{\gamma}
\def\d{\delta}

\def\vp{\varphi}
\def\vt{\vartheta}

\journal{Physics Letters B}

\begin{document} 
\begin{frontmatter}
  \title{Extended Einstein-Cartan theory \`a la Diakonov:\\ the field
    equations}

\author{Yuri N. Obukhov}
\address{Institute for Theoretical Physics, University of Cologne, 
50923 K\"oln, Germany }
\address{Dept.\ of Theoretical Physics, Moscow State
University, 117234 Moscow, Russia}

\author{Friedrich W. Hehl}
\address{Institute for Theoretical Physics, University of Cologne, 
50923 K\"oln, Germany}
\address{Dept.\ of Physics and Astronomy, University 
of Missouri, Columbia, MO 65211, USA}

\date{file ``dia11.tex'', 27 Feb 2012}

\begin {abstract}
  Diakonov formulated a model of a primordial Dirac spinor field
  interacting gravitationally within the geometric framework of the
  Poincar\'e gauge theory (PGT). Thus, the gravitational field
  variables are the orthonormal coframe (tetrad) and the Lorentz
  connection. A simple gravitational gauge Lagrangian is the
  Einstein-Cartan choice proportional to the curvature scalar plus a
  cosmological term. In Diakonov's model the coframe is eliminated by
  expressing it in terms of the primordial spinor. We derive the
  corresponding field equations for the first time. We extend the
  Diakonov model by additionally {\it eliminating the Lorentz
    connection,} but keeping local Lorentz covariance intact. Then, if
  we drop the Einstein-Cartan term in the Lagrangian, a nonlinear
  Heisenberg type spinor equation is recovered in the lowest
  approximation.
\end{abstract}


\begin{keyword}
  Poincar\'e gauge theory, Diakonov model, Einstein--Cartan theory,
  Heisenberg spinor equation\hfill {{\it file dia11.tex,
      27 Feb 2012}}
\end{keyword}
\end{frontmatter}


\section{Introduction}

Recently Diakonov \cite{dd} developed a scheme for the quantization of
the gravitational field interacting with a primordial {\it Dirac
  spinor} field $\psi$. His starting point is the framework of the
Poincar\'e gauge theory of gravity (PGT); therein gravity is described
by two independent fields, by the {\it orthonormal coframe}
$\vt^\a=e_i{}^\a dx^i$ (translational potential) and by the {\it
  Lorentz connection} $\Gamma^{\a\b}= \Gamma_i{}^{\a\b} dx^i=
-\Gamma^{\b\a}$ (Lorentz potential); for references see
\cite{Milutin,mag,Obukhov:2006gea}, our {\it notation} is explained at the end of
this introduction. Accordingly, the original field variables of
Diakonov's theory are $(\vt^\a,\Gamma^{\a\b},\psi,\overline{\psi})$.

In the Appendix we displayed the most general gravitational Lagrangian
of PGT that is {\it quadratic} in the torsion $T^\a$ and in the
curvature $R^{\a\b}=-R^{\b\a}$ and encompasses all parity even and odd
pieces
\cite{Obukhov:1987tz,Baekler:2010fr,Diakonov:2011fs,Baekler:2011jt}. At
first Diakonov considers only the simplest model, namely the first two
pieces on the right-hand-side of Eq.~(\ref{QMA}). They constitute the
gravitational Lagrangian of the Einstein-Cartan(-Sciama-Kibble) theory
(ECT), see \cite{Trautman}:
\begin{equation}\label{Lagrangian}
  V= -\,\frac{1}{2\kappa}\left(\eta_{\a\b}\wedge 
R^{\a\b}-2\Lambda\eta\right)\,.
\end{equation}
Here $\Lambda$ is the cosmological constant.  The matter part is
typically the minimally coupled Dirac Lagrangian 4-form
\begin{eqnarray}
  L_{\rm D} &=& 
  \frac{i}{2}\left(\overline{\psi}\;{}^\star\gamma\wedge D\psi
    +\overline{D\psi}\wedge{}^\star\gamma\,\psi\right)+{}^\star mc\,
  \overline{\psi}\psi\cr
  &=& 
  -\Big[\frac{i}{2}\, e_\a\rfloor\underbrace{\left(\overline{\psi}
      \gamma^\a D\psi-D\overline{\psi}\gamma^\a\,
      \psi\right)}_{\rm{1-form}} - \,mc\,
  \overline{\psi}\psi\Big]\eta.\label{10-4.10}
\end{eqnarray}
For later use we isolated in the second line the volume 4-form $\eta$. 

The Einstein-Cartan-Dirac theory of gravity has the total Lagrangian
$L= V+L_{\rm D}$, with the action $W=\int \!\! L$. The two gravitational field 
equations read
\begin{eqnarray}
{\frac 1{2\kappa}}\,R^{\rho\sigma}\wedge\eta_{\alpha\rho\sigma} -
{\frac \Lambda\kappa}\,\eta_\alpha &=& \Sigma^{\rm D}_\alpha = 
{\frac i2}\left(\overline{\psi}\,{}^\star\!\gamma D_\alpha\psi -
D_\alpha\overline{\psi}\,{}^\star\!\gamma\psi\right),\label{D1}\\
{\frac 1{2\kappa}}\,T^{\rho}\wedge\eta_{\alpha\beta\rho} &=& \tau^{\rm  D}_{\alpha\beta} 
=  -\,{\frac 18}\,\overline{\psi}({}^\star\gamma\sigma_{\alpha\beta}
+ \sigma_{\alpha\beta}{}^\star\gamma)\psi.\label{D2}
\end{eqnarray}
As sources of the field equations act the canonical 3-forms of
energy-momentum $\Sigma^{\rm D}_\alpha$ and spin angular momentum and
$ \tau^{\rm D}_{\alpha\beta}$ of the Dirac field, respectively, see
\cite{Harvey}. The second field equation (\ref{D2}) is algebraic in
the torsion and can be resolved with respect to the {\it contortion}
1-form $K_\alpha{}^\beta = \widetilde{\Gamma}_\alpha{}^\beta -
\Gamma_\alpha {}^\beta$, where
$\widetilde{\Gamma}_\alpha{}^\beta(\vt^\vt,d\vt^\a)$ is the Riemannian
piece of the connection and $T^\alpha =
K^\alpha{}_\beta\wedge\vartheta^\beta$:
\begin{eqnarray}
K^{\alpha\beta} = -\,\kappa\,{}^\star\!\tau^{\rm D}{}^{\alpha\beta}
= {\frac \kappa 4}\,\eta^{\alpha\beta\rho}\,\overline{\psi}\gamma_\rho
\gamma_5\psi = {\frac \kappa 4}\,{}^\star(\vartheta^{\alpha\beta}\wedge
\overline{\psi}\gamma\gamma_5\psi).\label{KT}
\end{eqnarray}
In vacuum, where the Dirac field vanishes, contortion and torsion
vanish, too, and we recover the vacuum Einstein theory of gravity.
\medskip

\begin{footnotesize}
  \noindent{\bf Notation} \cite{mag,Obukhov:2006gea}: Latin letters
  $i,j,k,...=0,1,2,3$ denote (holonomic) coordinate indices, Greek
  letters $\a,\b,\g,...=\hat{0},\hat{1},\hat{2},\hat{3}$ (anholonomic)
  frame indices. Symmetrization over indices is denoted by
  $(\a\b):=\{\a\b+\b\a\}/2$, antisymmetrization by
  $[\a\b]:=\{\a\b-\b\a\}/2$. The frame $e_\b=e^j{}_\b\partial_j$ is
  dual to the coframe: $e_\b\rfloor\vt^\a=\d^\a_\b$. Greek indices are
  raised and lowered by means of the Minkowski metric $g_{\a\b}={\rm
    diag}(1,-1,-1,-1)$. The volume 4-form is denoted by
  $\eta={}^\star1$, and $\eta_\a={}^\star \vt_\a,\;
  \eta_{\a\b}={}^\star \vt_{\a\b},\; \eta_{\a\b\g}={}^\star
  \vt_{\a\b\g},\; \eta_{\a\b\g\d}={}^\star \vt_{\a\b\g\d}$, where
  $^\star$ is the Hodge star operator and $\vt^{\a\b}:=\vt^\a\wedge
  \vt^\b,\;{\rm etc}$. Furthermore, $ T^\a:=D\vt^\a=T_{ij}{}^\a
  dx^i\wedge dx^j/2\,,\> T_{ij}{}^\a =
  2(\partial_{[i}\vt_{j]}{}^{\a}+\Gamma_{[ij]}{}^\a )\,,\>
  R^{\a\b}:=d\Gamma^{\a\b}+\Gamma^{\a\g}\wedge
  \Gamma^\b{}_\g=R_{ij}{}^{\a\b}dx^i\wedge dx^j/2\,,\>
  R_{ij}{}^{\a\b}=2(\partial_{[i}\Gamma_{j]}{}^{\a\b}
  +\Gamma_{[i}{}^{\a\g}\Gamma_{j]\,\cdot\,\g}^{\hspace{5pt}\b})\,,\>
  {\rm Ric}_\a:= e_\b\rfloor R_\a{}^\b={\rm Ric}_{\b\a}\,\vt^\b,\,{\rm
    with}\, {\rm Ric}_{\a\b}=R_{\g\a\b}{}^\g,$ $R:={\rm
    Ric}_\a{}^\a=R_{\b\a}{}^{\a\b}.$ Our master formula for the
  variation of coframe and metric is \cite{Muench:1998ay}:
\begin{eqnarray}
  \left(\delta \ensuremath{{}^\star}
        -\ensuremath{{}^\star} \delta\right) \phi = \;\delta
      \vartheta^\alpha \wedge \left(e_\alpha \rfloor
        \ensuremath{{}^\star} \phi \right) - \ensuremath{{}^\star}
      \big[ \delta \vartheta^\alpha \wedge \left( e_\alpha \rfloor
        \phi \right)\big]  + \,\delta
      g_{\alpha\beta}\Big[\vartheta^{(\alpha}\! \wedge\!
      ( e^{\beta)}
 \rfloor \ensuremath{{}^\star} \phi ) - \frac{1}{2}\,g^{\alpha\beta}\;
      \ensuremath{{}^\star} \phi\Big] \;.\label{eq:var-hodge}
\end{eqnarray}
In the Dirac theory we have
  the 1-form $\gamma:=\gamma_\a\vt^\a$ and for the Dirac adjoint
  $\overline{\psi} := \psi^\dagger\gamma^0$; the exterior spinor
  covariant derivative is given by $ D = d
  +{ { i}}\sigma_{\alpha\beta}\Gamma^{\alpha\beta}/4$ and the Lorentz
  algebra generators by $\sigma^{\alpha\beta} = i\gamma^{[\alpha}
  \gamma^{\beta]} $. Moreover, we put $\hbar=1, c=1$.
\end{footnotesize}
\section{Diakonov's model}

Diakonov \cite{dd} proposed recently a model of ``microscopic quantum
gravity" such that the coframe $\vt^\a$ is dropped as an independent
field and expressed in terms of a primordial anticommuting Dirac field
$\Psi$, that is $\vt^\a=\vt^\a(\Psi,\overline{\Psi})$. Following
similar ideas of Akama \cite{Akama:1978pg}, he made the
ansatz 
\begin{equation}
 {\ell^{-4}} \vartheta^\alpha = {\frac i2}\left(\overline{\Psi}
    \gamma^\alpha D\Psi
    - D\overline{\Psi}\gamma^\alpha\Psi\right)\label{vpsi}
\end{equation}
or $ \varphi^\a=0$, with 
\begin{equation}
  \varphi^\a(\vt^a,\Gamma^{\a\b}, \Psi,\overline{\Psi},d\Psi,d\overline{\Psi}):=
  {\frac  i2}\left(  \overline{\Psi}\gamma^\alpha D\Psi - D\overline{\Psi}
    \gamma^\alpha\Psi\right)- {\ell^{-4}} \vartheta^\alpha.\label{vpsi*}
\end{equation}
The constant $\ell$ carries the dimension of a length. Apparently, the
primordial spinor is assumed to be massless. In contrast to $\vt^\a$,
the connection $\Gamma^{\a\b}$ is upheld as an independent field.

We recognize on the right-hand-side of (\ref{vpsi}) the 1-form marked
in (\ref{10-4.10}). A comparison with $ \Sigma^{\rm D}_\alpha$ in
(\ref{D1}) and some algebra yields an alternative form of
(\ref{vpsi}). We first define the {\it transposed} of an arbitrary
covector-valued 1-form $E_\a=E_{i\a}dx^i$ as $\stackrel{\frown}{E_\a}
:= E_\a +e_\a \rfloor(E_\b\wedge\vt^\b)=(e^\a\rfloor E_\b)\vt^\b$. It
is simple to verify that 
$\stackrel{\frown}{E_{i\a}}=E_{\a i}$. With this new
transposition operator we can write Diakonov's formula (\ref{vpsi})
succintly as
\begin{equation}\label{Hooke}
  \Sigma_\a^{\rm D}=  {\ell^{-4}}g_{\a\b}\,^\star\!\! \stackrel{\frown}{\vt^\b}
\end{equation}
This is reminiscent of the Hooke's type constitutive law in Cosserat
elasticity, see \cite{Hehl:2007bn}: the (asymmetric) force stress
3-form $\Sigma^{\rm D}$ is proportional to the distortion 1-form
$\vt$, with $ {\ell^{-4}}g_{\a\b}$ as the elasticity modulus.\footnote{This
  similarity would be even more pronounced, if one dropped the
  transposition operator and took the gauge-theoretically more
  satisfactory ansatz $\Sigma^{\rm D}_\a=\ell^{-4}\,^\star\!\vt_\a$.}

Diakonov \cite{dd} assumes that the primordial spinor matter acts only
via the coframe $\vt^\a$ according to (\ref{vpsi}), (\ref{vpsi*}), or
(\ref{Hooke}). Thus, no explicit matter part enters the
Lagrangian. Most conveniently we stick with the original field
variables $(\vt^\a,\Gamma^{\a\b},\Psi,\overline{\Psi})$ and add the
constraint $\vp^\a=0$ with suitable Lagrange multiplier 3-forms
$\lambda_\a$ to the EC-Lagrangian $V$. In this way the local
Poincar\'e covariance of the PGT is upheld. Thus, the {\it Diakonov
  Lagrangian} reads
\begin{eqnarray}
  L =V+L_{\rm{mat}}
  = -\,{\frac 1{2\kappa}}\left(a_0R^{\alpha\beta}\wedge\eta_{\alpha\beta} 
    - 2\Lambda\,\eta\right)+L_{\rm{mat}}\,,\label{L}
\end{eqnarray}
with the matter Lagrangian
\begin{equation}\label{Lmatter}
  L_{\rm{mat}}= L(\vartheta^\alpha, \Gamma^{\a\b},\Psi,\overline{\Psi}, 
  d\Psi, d\overline{\Psi},
  \lambda_\alpha 
)  :=\vp^\a\wedge\lambda_\a.
\end{equation}
The dimensionless constant parameter $a_0$ is introduced for
generality, compare (\ref{QMA}). We can immediately calculate the
translation and Lorentz {\it excitations}:
\begin{equation}\label{excit}
H_\a:=-\,\frac{\partial V}{\partial T^\a}=0,\quad
H_{\a\b}:=-\,\frac{\partial V}{ \partial R^{\a\b}}=\frac{a_0}{2\kappa}\,\eta_{\a\b}.
\end{equation}
In turn, we find for the energy-momentum and the spin angular momentum
of the {\it gauge fields}
\begin{eqnarray}
  E_\alpha &:=& e_\alpha\rfloor V + (e_\a\rfloor T^\b)\wedge H_\b
  +(e_\alpha\rfloor R^{\rho\sigma})\wedge H_{\rho\sigma}\nonumber\\ &=&
  -\,{\frac {a_0}{2\kappa}}\,R^{\rho\sigma}\wedge\eta_{\alpha\rho\sigma} +
  {\frac \Lambda\kappa}\,\eta_\alpha,\label{Ea}\\
  E_{\alpha\beta} &:=&-\vt_{[\a}\wedge H_{\b]} =0.\label{Eab}
\end{eqnarray}
Concentrating now on the matter Lagrangian (\ref{Lmatter}), we find
for the material energy-momentum and spin currents, respectively,
\begin{eqnarray}\label{EMmatter}
  \Sigma_\a&:=&\frac{\d L_{\rm{mat}}}{\d\vt^a}= -\, {\ell^{-4}}\lambda_\a
,\\
      \tau_{\a\b}&:=&\frac{\d L_{\rm{mat}}}{\d\Gamma^{\a\b}}=
 -\,{\frac 18}\,\overline{\Psi}(\gamma^\rho\sigma_{\alpha\beta}
+ \sigma_{\alpha\beta}\gamma^\rho)\Psi\,\lambda_\rho,\label{tauab}
\end{eqnarray}


\section{Field equations}

The two classical field equations of gravity read
\begin{eqnarray}
DH_\alpha - E_\alpha &=& \Sigma_\alpha,\label{1st}\\
DH_{\alpha\beta} - E_{\alpha\beta} &=& \tau_{\alpha\beta}.\label{2nd}
\end{eqnarray}
Consequently, the {\it first} field equation (\ref{1st}), with the
help of (\ref{excit})$_1$ and (\ref{Ea}), determines the Lagrange
multiplier
\begin{equation}
  {\ell^{-4}}\lambda_\alpha = -\,{\frac {a_0}{2\kappa}}\,
  R^{\rho\sigma}\wedge\eta_{\alpha\rho\sigma}
  + {\frac \Lambda\kappa}\,\eta_\alpha,\label{La}
\end{equation}
whereas the {\it second} field equation (\ref{2nd}), by means of
(\ref{excit})$_2$ and (\ref{Eab}), reduces to a nontrivial equation
for the torsion:
\begin{equation}
{\frac {a_0}{2\kappa}}\,T^\rho\wedge\eta_{\alpha\beta\rho} = 
-\,{\frac 18}\,\overline{\Psi}(\gamma^\rho\sigma_{\alpha\beta}
+ \sigma_{\alpha\beta}\gamma^\rho)\Psi\,\lambda_\rho.\label{Ta}
\end{equation}
If we substitute the Lagrange multiplier $\lambda_\a$ into the
right-hand-side of (\ref{Ta}), we recognize that, in contrast to the
ECT, in Diakonov's theory the {\it torsion is non-trivial.}

Eventually, variation of (\ref{L}) with respect to the primordial
spinor yields the {\it generalized Dirac} equation
\begin{equation}
i\gamma^\alpha \lambda_\alpha\wedge D\Psi - {\frac i2}\gamma^\alpha(D\lambda_\alpha)
\,\Psi = 0.\label{De}
\end{equation}
Using (\ref{La}), we have for the derivative of the Lagrange multiplier
\begin{equation}
  D\lambda_\alpha = -\,\frac{ \ell^{4}}{\kappa}
  \left({\frac{a_0}2}\,\eta_{\alpha\beta\rho\sigma}
    R^{\rho\sigma} - \Lambda\eta_{\alpha\beta}\right)\wedge T^\beta .\label{DLa}
\end{equation}

Variation with respect to the Lagrange multiplier 3-form
$\lambda_\alpha$ yields the {\it constraint}
\begin{equation}
\varphi^\alpha = 0 \label{va}
\end{equation}
which reproduces (\ref{vpsi}) or the ``constitutive law''
(\ref{Hooke}).

\section{A possible generalization of Diakonov's model?}

Since Diakonov wanted to keep the local Lorentz invariance of the
action of his model, he did not express the Lorentz connection in
terms of the primordial spinor. But there is no convincing rationale
behind this argument. With our interpretation (\ref{Hooke}) of the
Akama--Diakonov ansatz (\ref{vpsi}), the generalization to the
elimination of the Lorentz connection is straightforward. In Cosserat
elasticity, the second constitutive law relates the spin moment stress
3-form (torque) $\tau^{\rm D}$ linearly to the rotational deformation measure,
the contortion 1-form $K$ (see \cite{Hehl:2007bn}, Sec.\ 3):
\begin{equation}\label{Cosserat}
  \tau^{{\rm D}}_{\a\b}={{\rm L}^{-2}}g_{\a\g}g_{\b\d}\, ^{\star}\!
  K^{\g\d}={{\rm L}^{-2}}\,
^\star\!\!\left(\widetilde{\Gamma}_{\a\b}(\vt,d\vt)-\Gamma_{\a\b} \right)\,.
\end{equation}
Here the constant $\rm L$ of the ``rotational modulus'' ${\rm
  L}^{-2}g_{\a\g}g_{\b\d}$ carries the dimension of a length. As new
constraint we have then $ \vp^{\a\b}=0$, with
\begin{equation}\label{constraint2}
  \vp^{\a\b}(\vt^a,d\vt^a,\Gamma^{\a\b},\Psi,\overline{\Psi}):=
  \frac 14 \,^\star\!\!
  \left(\vt^{\a\b}\wedge\overline{\Psi}\gamma\gamma_5\Psi \right) -
  {{\rm L}^{-2}} \left(\widetilde{\Gamma}^{\a\b}(\vt,d\vt)
    -\Gamma^{\a\b} \right)\,
\end{equation}
[see \cite{Harvey}, Eq.~(22)]. As a result, the matter Lagrangian
(\ref{Lmatter}) is generalized to
\begin{equation}\label{Lmatter2}
  L_{\rm{mat}}= L(\vartheta^\alpha,d\vt^\a,\Gamma^{\a\b}, \Psi,\overline{\Psi}, 
  d\Psi, d\overline{\Psi}, \lambda_\alpha,\lambda_{\a\b})
    = \vp^\a\wedge\lambda_\a + \varphi^{\alpha\beta}\wedge\lambda_{\a\b},
\end{equation}
where we introduced a second Lagrange multiplier 3-form
$\lambda_{\a\b}=-\lambda_{\b\a}$.  

We could now execute the variational calculations
explicitly.\footnote{Direct variation is spelled out in the following
  formula:
\begin{eqnarray}\label{EMmatter2x}\nonumber
  \Sigma_\a&=& -\,{\ell^{-4}}\lambda_\a
  +\left({\d\left[^\star\!\!
        \left(\vt^{\mu\nu}\wedge\overline{\Psi}\gamma
          \gamma_5\Psi \right)/4 -
        {{\rm L}^{-2}} \widetilde{\Gamma}^{\mu\nu}(\vt,d\vt)
      \right]}/{\d\vt^\a}\right)\wedge\lambda_{\mu\nu}\,.
\end{eqnarray}
In this case, the master formula (\ref{eq:var-hodge}) has to be
applied repeatedly. In a similar way, we can also compute
$\tau_{\a\b}$.} However, there is a simpler method applicable. We
consider the Lagrange multiplier as an additional matter field and
employ the Lagrange-Noether machinery as described in \cite{mag},
Sec.\ 5, for example. The constraints $\vp^\a=0$ and $\vp^{\a\b}=0$
are used thereby. After some algebra, we find that for the matter
Lagrangian (\ref{Lmatter2}) the formulas (\ref{EMmatter}) and
(\ref{tauab}) are generalized as follows:
\begin{eqnarray}\label{Sigmaa2}
  \Sigma_\a&:=&\frac{\d L_{\rm{mat}}}{\d\vt^a}= -\,{\ell^{-4}}\lambda_\a
  +  {{\rm L}^{-2}}\stackrel{\frown}{D}\xi_\alpha
  ,\\
  \tau_{\a\b}&:=&\frac{\d L_{\rm{mat}}}{\d\Gamma^{\a\b}}=
  -\,{\frac 18}\,\overline{\Psi}(\gamma^\rho\sigma_{\alpha\beta}
  + \sigma_{\alpha\beta}\gamma^\rho)\Psi\,\lambda_\rho +
  {{\rm L}^{-2}}  \lambda_{\a\b}.\label{tauab2}
\end{eqnarray}
Here we introduced the
covector-valued 2-form $\xi_\a$ which is equivalent to the
antisymmetric tensor-valued Lagrange multiplier 3-form $\lambda_{\a\b}$:
\begin{equation}
\xi_\alpha ={{\rm L}^{2}} {\frac {\partial L_{\rm{mat}}}{\partial T^\alpha}} = 
2e^\rho\rfloor\lambda_{\rho\alpha} + {\frac 12}\,\vartheta_\alpha\wedge
(e^\rho\rfloor e^\sigma\rfloor\lambda_{\rho\sigma})\,;\label{xia}
\end{equation}
both fields have 24 independent components (see \cite{mag}, Eq.\
(5.1.24)). Furthermore, $\stackrel{\frown}{D}\xi_\alpha$ is the
covariant exterior derivative with respect to the transposed
connection $\stackrel{\frown}{\Gamma}_\a{}^{\!\!\b}:=
{\Gamma}_\a{}^\b-e_\a\rfloor T^\b$, see \cite{mag}, Eq.\ (3.11.9).

Let us collect our results. We have a Lagrangian $L=V+L_{\rm mat}$,
with the matter Lagrangian (\ref{Lmatter2}). In the framework of the
Poincar\'e gauge theory (PGT), $V$ is the quadratic gauge Lagrangian
(\ref{QMA}). The general field equations are (\ref{1st}) and
(\ref{2nd}). If we substitute the excitations $(H_\a,H_{\a\b})$, see
(\ref{excit}), and the gauge currents $(E_\a,E_{\a\b})$, see
(\ref{Ea}) and (\ref{Eab}), respectively, we find
\begin{eqnarray}\label{FIRST}
  -D\frac{\partial V}{\partial T^\a}-e_\a\rfloor V
  +(e_\a\rfloor T^\b)\wedge\frac{\partial V}{\partial T^\b}
  +(e_\a\rfloor R^{\b\g})\wedge\frac{\partial V}{\partial R^{\b\g}}
  &=&\Sigma_\a\,,\\
  -D\frac{\partial V}{\partial R^{\a\b}}
  +\vt_{[\a}\wedge\frac{\partial V}{\partial T^{\b]}} &=&\tau_{\a\b}\,.
\label{SECOND}
\end{eqnarray}
The sources on the right-hand-sides of these gauge field equations are
provided by (\ref{Sigmaa2}) and (\ref{tauab2}). In other words, Eqs.\
(\ref{FIRST}) and (\ref{SECOND}) merely determine the Lagrange
multipliers $\lambda_\a$ and $\lambda_{\a\b}$; and they do it for the
complete quadratic Lagrangian displayed in (\ref{QMA}). In other
words, we have eliminated by this procedure coframe and Lorentz
connection altogether within the framework of the quadratic PGT.

For the general Lagrangian (\ref{QMA}), these field equations are very
complicated. For the Diakonov Lagrangian with the Einstein-Cartan term
plus cosmological constant, see (\ref{L}), they reduce in our
generalized framework to
\begin{eqnarray}\label{dia1}
{\frac {a_0}{2\kappa}}\,R^{\rho\sigma}\wedge\eta_{\alpha\rho\sigma} -
{\frac \Lambda\kappa}\,\eta_\alpha &=&  -\,{\ell^{-4}}\lambda_\a
+  {{\rm L}^{-2}}\stackrel{\frown}{D}\xi_\alpha
,\\
{\frac {a_0}{2\kappa}}\,T^{\rho}\wedge\eta_{\alpha\beta\rho} &=& 
-\,{\frac 18}\,\overline{\Psi}(\gamma^\rho\sigma_{\alpha\beta}
+ \sigma_{\alpha\beta}\gamma^\rho)\Psi\,\lambda_\rho +
{{\rm L}^{-2}}  \lambda_{\a\b}.
\label{dia2}
\end{eqnarray}
Note that for ${\rm L}\rightarrow \infty$, we {\it recover} the
Diakonov model, see (\ref{La}) and (\ref{Ta}). The dynamics of the
theory is contained in the nonlinear spinor equation that we obtain by
varying the action with respect to the spinor field:
\begin{equation}
i\gamma^\alpha \lambda_\alpha\wedge D\Psi - {\frac i2}\gamma^\alpha(D\lambda_\alpha)
\,\Psi + {\frac 14}\,\vartheta^{\a\b}\wedge{}^\star\lambda_{\a\b}\wedge
\gamma\gamma_5\Psi = 0.\label{De2}
\end{equation}
Accordingly, our generalized Diakonov model, in which eventually only
the primordial spinor is left as field variable, is controlled by the
field equations (\ref{dia1}), (\ref{dia2}), (\ref{De2}),
$\varphi^\a=0$, $\varphi^{\a\b}=0$.

As already observed by Akama \cite{Akama:1978pg} and Diakonov
\cite{dd}, the absolute simplest model would be to {\it drop the
  Einstein-Cartan Lagrangain} by putting $a_0$ to zero.  This reduces
the gravitational Lagrangian to the cosmological term, that is, just
to a volume integral:
\begin{equation}
L = {\frac {\Lambda}{\kappa}}\,\eta + \vp^\a\wedge\lambda_\a 
+ \varphi^{\alpha\beta}\wedge\lambda_{\a\b}.\label{Lcomb}
\end{equation}
Then the field equations (\ref{dia1}), (\ref{dia2}), with $a_0=0$,
determine the Lagrange multipliers according to
\begin{eqnarray}
\ell^{-4}  \lambda_\a &=& {\frac {\Lambda}{\kappa}}\,\eta_\a 
  +\frac{1}{{\rm L}^{2}} \stackrel{\frown}{D}\xi_\alpha 
,\label{l1}\\
 {\rm L}^{-2} \lambda_{\a\b} &=& \frac{1}{8}\,\overline{\Psi}(
  \gamma^\rho\sigma_{\alpha\beta}
  + \sigma_{\alpha\beta}\gamma^\rho)\Psi\,\lambda_\rho.\label{l2}
\end{eqnarray}

In general, the differential-algebraic system (\ref{l1}), (\ref{l2}),
(\ref{De2}) is quite nontrivial. One can try to {\it solve it
  iteratively,} then in the first approximation we find
\begin{equation}
  \lambda_\a = {\frac {\Lambda\ell^{4}}{\kappa}}\,\eta_\a,
  \qquad {}^\star\lambda_{\a\b} = -\,{\frac {\Lambda\ell^{4}
      {\rm L}^{2}}{4\kappa}}\,\eta_{\a\b\rho\sigma}\vartheta^\rho
  \,\overline\Psi\gamma^\sigma\gamma_5\Psi.\label{approx}
\end{equation}
As a consequence (\ref{De2}) reduces to
\begin{equation} {}^\star\gamma\wedge iD\Psi - {\frac
    38}\,{\rm L}^2\,^\star(\overline\Psi\gamma\gamma_5\Psi)
  \wedge\gamma\gamma_5\Psi = 0.\label{nonlinD}
\end{equation}
In this zeroth approximation, the gravitational constant
$\kappa=\ell_{{\rm Pl}}^2\,$, the cosmological constant
$\Lambda=1/\ell_{\rm cos}^2\,$, and the constant $\ell$ all drop
out. In the Einstein-Cartan theory we have a critical length
$\ell_{\rm EC}=(\lambda_{\rm C}\ell_{\rm Pl}^2)^{1/3}$, where
$\lambda_{\rm C}$ is the Compton wavelength typically of the
nucleon. The problem of these different length scales was not
discussed by Diakonov. One guess would be that $\rm L$ in
(\ref{nonlinD}) has to be identified with $\ell_{\rm EC}$, but the
situation is far from clear to us. In any case, the dimensionless
constant $\beta:=({\Lambda\ell^4}/{\kappa})^{1/2}=\ell^2/(\ell_{\rm
  Pl}\ell_{\rm cos})$ in (\ref{approx})$_1$, that is, in
$\lambda_\a=\b^2\eta_\a$, possibly plays an important role.

Thus, in the lowest approximation we recover a nonlinear spinor
equation of the Heisenberg-Pauli type \cite{HeisenbergPauli}, which
was once part of one of the most advanced models \cite{Heisenberg}
that attempted to describe all physical interactions in terms of a
fundamental fermion field, see also Ivanenko's work on nonlinear
spinor equations \cite{Ivanenko}. Kibble \cite{Kibble} and Rodichev
\cite{Rodichev} were the first to recognize that such a type of
equation emerges in the context of the Dirac theory automatically
provided spacetime carries {\it Cartan's torsion} in the context of a
gravitational gauge theory.

\section{Discussion}

Let us consider a special case of our generalized model that is in a
sense {\it complementary} to the model of Diakonov. Namely, we impose
only the second constraint (\ref{constraint2}) for the connection, but
forget about the original constraint (\ref{vpsi*}) for the
coframe. Then the gravitational field equations
(\ref{1st}), (\ref{2nd}) reduce to
\begin{eqnarray}
{\frac {a_0}{2\kappa}}\,R^{\rho\sigma}\wedge\eta_{\alpha\rho\sigma} + {\frac \Lambda 
\kappa}\,\eta_\alpha &=& {\frac {a_0}{2\kappa}}\left[T_\alpha\wedge\overline{\Psi}
\gamma\gamma_5\Psi + \vartheta_\alpha\wedge D(\overline{\Psi}
\gamma\gamma_5\Psi)\right],\label{1st2}\\
{\rm L}^{-2}\lambda_{\a\b} &=& {\frac {a_0}{2\kappa}}\,T^\rho\wedge\eta_{\a\b\rho} =
-\,{\frac {a_0}{4\kappa}}\,\vartheta_{\a\b}\wedge\overline{\Psi}
\gamma\gamma_5\Psi.\label{2nd2}
\end{eqnarray}
We thus see that in the complementary picture with only a connection
constraint, the coframe is determined from the Einstein like
gravitational field equation (\ref{1st2}) with some effective
energy-momentum current constructed from the spinor fields. The second
field equation (\ref{2nd2}) determines the Lagrange multiplier
$\lambda_{\a\b}$.  This is very different from the original Diakonov
model with the coframe constraint. The first equation (\ref{La})
determines the Lagrange multiplier $\lambda_\a$, whereas the second
equation (\ref{Ta}) describes the nontrivial spacetime torsion.

In this sense, the Diakonov theory can be called a
``Cartan-connection'' model, and the complementary theory an
``Einstein-coframe'' (or ``-tetrad'') model. Diakonov uses a Hooke's type
law as constraint
\begin{equation}\label{Hooke's}
  \Sigma_\a^{\rm D}=  {\ell^{-4}}g_{\a\b}\,^\star\!\! \stackrel{\frown}{\vt^\b},
\end{equation}
in the complementary case a MacCullagh's type law, see
\cite{Darrigol}, for the rotationally elastic aether is employed (see
its modern incarnation in \cite{Christian+Yuri} and compare
\cite{Lazar:2010,Lazar:2009}),
\begin{equation}\label{MacCullagh's}
  \tau^{\rm D}_{\a\b}={\rm L}^{-2} g_{\a\g}g_{\b\d}\,^\star\!{K^{\g\d}}\,.
\end{equation}
In our extended Diakonov model we postulate the constitutive equations
of an elastic Cosserat continuum, which is responsive to translational
and rotational deformations, that is, we take both, Hooke's and
MacCullagh's law at the same time. Such pictures from continuum
mechanics (classical field theory) helped also Hammad
\cite{Hammad:2012ki} to find an entropy functional for a
graviatational theory acting in a Riemann-Cartan spacetime.

The extended Diakonov model appears to be a generalization of the old
Heisenberg nonlinear spinor theory that is recovered in the lowest
approximation in equation (\ref{nonlinD}). The quantization of the
full theory is a difficult task, and it will not be discussed
here. Diakonov \cite{dd} is using a lattice approach and work is in
progress.

Despite the fact that the ansatz for the coframe (\ref{vpsi}) was
inspired by the work of Akama \cite{Akama:1978pg}, the model of Akama
is only invariant under the global Lorentz group. In contrast, the
Diakonov \cite{dd} and the extended Diakonov models are explicitly
invariant under {\it local} Lorentz transformations. This is a result
of the consistent use of the geometrical framework of the Poincar\'e
gauge theory of gravity.

\section*{Appendix: Most general quadratic Lagrangian in PGT}

The PGT with its two field equations (\ref{1st}) and (\ref{2nd}) is
only complete if we specify its gauge Lagrangian. As a typical gauge
theory, the Lagrangian is quadratic in the field strengths torsion
$T^\a={\scriptstyle{\sum_{I=1,2,3}}}{}^{(I)}T^\a$ and curvature
$R^{\a\b}={\scriptstyle{\sum_{K=1,2,...,6}}}{}^{(K)}R^{\a\b}$; here
torsion and curvature are represented as sums in terms of their
irreducible pieces.

If we introduce the notations $R$ and $X$ for the curvature scalar and
the curvature pseudoscalar, then we find $^{(6)}R_{\a\b} =
-R\,\vt_{\a\b}/12$ and $^{(3)}R_{\a\b}= -X\eta_{\a\b}/12$,
respectively; moreover, for the torsion we can define the 1-forms of
$\cal A$ and $\cal V$ for the axial vector and the vector torsion
$^{(3)}T^{\a} = {}^{\star}({\cal A}\wedge\, {\vartheta}^{\a})/3$ and
$^{(2)}T^\a=-({\cal V}\wedge\vt^\a)/3$, respectively. Our final
gravitational Lagrangian is then
\cite{Obukhov:1987tz,Baekler:2010fr,Diakonov:2011fs,Baekler:2011jt}
\begin{eqnarray} \label{QMA}\nonumber V =
  -\,\frac{1}{2\kappa}[\,\left(\,a_0R-2\Lambda+{b_0}X)\,\eta
\right. \hspace{150pt}\\
  \left. +\frac{a_{2}}{3} {\cal V}\wedge {}^{\star\!} {\cal V}
    -\frac{a_{3}}{3}{{\cal A} \wedge{} ^{\star\!\!\!}{\cal A}}
    -\frac{2{\sigma}_{2}}{3}{ {\cal V}\wedge{} ^{\star\!\!\!} {\cal
        A}}+ a_{1}{}^{(1)}T^\alpha \wedge
    {}^{\star(1)}T_\a \right]\nonumber \\
  -\frac{1}{2\varrho} \left[ (\frac{w_6}{12} R^2- \frac{w_3}{12} {X^2}
    + \frac{\mu_3}{12} R X)\,\eta + w_{4}{}^{(4)}\!R^{\a\b}\wedge
    {}^{\star(4)}\!R_{\a\b}\right.\hspace{20pt}\nonumber\cr  \\
  \left. +{}^{(2)}\!R^{\a\b}\wedge ( w_2{}^{\star(2)}\!R_{\a\b}
    +\mu_2{}^{(4)}\!R_{\a\b})+{}^{(5)}\!R^{\a\b}
    \wedge(w_5{}^{\star(5)}\!R_{\a\b}
    +\mu_4{}^{(5)}\!R_{\a\b})\right].
\end{eqnarray}
The first two lines represent weak gravity with the gravitational
constant $\kappa$, the last two lines strong gravity with the
dimensionless coupling constant $\varrho$. The parity odd pieces are
those with the constants $b_0,\sigma_2;\mu_2,\mu_3,\mu_4$. In a
Riemann space (where $X=0$), only two terms of the first line and
likewise two terms in the third line survive. All these 4 terms are
parity even, that is, only torsion brings in parity odd pieces into
the gravitational Lagrangian.

\section*{Acknowledgments}

This work was supported by the German-Israeli Foundation for
Scientific Research and Development (GIF), Research Grant No.\
1078-107.14/2009.  \bigskip



\end{document}